\documentclass[12pt,preprint]{aastex}
\usepackage{hyperref}
\pdfoutput=1

\usepackage{float}
\usepackage{graphicx}
\usepackage{epsfig}
\usepackage[perpage,symbol*]{footmisc}
\usepackage{natbib}
\usepackage{dcolumn}
\usepackage[figureright]{rotating}

\shorttitle{A 6.7 GHz Methanol Maser Survey}
\shortauthors{Yang et al.}

\begin{document}

\title{A 6.7 GHz Methanol Maser Survey at High Galactic Latitudes}

\author{Kai Yang\altaffilmark{1,2}, Xi Chen\altaffilmark{1,3,4}, Zhi-Qiang Shen\altaffilmark{1,4}, Xiao-Qiong Li\altaffilmark{1,2}, Jun-Zhi Wang\altaffilmark{1,4}, Dong-Rong Jiang\altaffilmark{1,4}, Juan Li\altaffilmark{1,4}, Jian Dong\altaffilmark{1,4}, Ya-Jun Wu\altaffilmark{1,4}, Hai-Hua Qiao\altaffilmark{1,2}, Zhiyuan Ren\altaffilmark{4,5}}

\altaffiltext{1}{Shanghai Astronomical Observatory, Chinese Academy of Sciences, 80 Nandan Road, Shanghai 200030, China; yangkai@shao.ac.cn, chenxi@shao.ac.cn, zshen@shao.ac.cn.}
\altaffiltext{2}{University of Chinese Academy of Sciences, 19A Yuquanlu, Beijing 100049, China.}
\altaffiltext{3}{Center for Astrophysics, Guangzhou University, Guangzhou 510006, China; chenxi@gzhu.edu.cn.}
\altaffiltext{4}{Key Laboratory of Radio Astronomy, Chinese Academy of Sciences, China.}
\altaffiltext{5}{National Astronomical Observatories, Chinese Academy of Sciences, A20 Datun Road, Chaoyang District, Beijing 100012, China.}

\begin{abstract}
 We performed a systematic 6.7 GHz Class \uppercase\expandafter{\romannumeral2} methanol maser survey using the Shanghai Tianma Radio Telescope toward targets selected from the all-sky \emph{Wide-Field Infrared Survey Explorer (WISE)} point catalog. In this paper, we report the results from the survey of those at high Galactic latitudes, i.e. $|b|>$ 2$^\circ$. Of 1473 selected \emph{WISE} point sources at high latitude, 17 point positions that were actually associated with 12 sources were detected with maser emission, reflecting the rarity (1$-$2\%) of methanol masers in the region away from the Galactic plane. Out of the 12 sources, 3 are detected for the first time. The spectral energy distribution (SED) at infrared bands shows that these new detected masers occur in the massive star forming regions. Compared to previous detections, the methanol maser changes significantly in both spectral profiles and flux densities. The infrared \emph{WISE} images show that almost all of these masers are located in the positions of the bright \emph{WISE} point sources. Compared with the methanol masers at the Galactic plane, these high-latitude methanol masers provide good tracers for investigating the physics and kinematics around massive young stellar objects, because they are believed to be less affected by the surrounding cluster environment.

\end{abstract}

\keywords{masers --- stars: formation --- ISM: molecules --- radio lines: ISM}

\section{Introduction}

 Methanol masers are very useful tools for investigating physics and kinematics in massive star forming regions. Methanol masers have been divided into two classes \citep[Class \uppercase\expandafter{\romannumeral1} and Class \uppercase\expandafter{\romannumeral2},][]{Bat1987}. \citet{Vor2010,Vor2014} concluded that Class \uppercase\expandafter{\romannumeral1} methanol masers are pumped by collision excitation in mildly shocked molecular gas and reside on scales of $\sim$1 pc in a single star-formation region. On the contrary, Class \uppercase\expandafter{\romannumeral2} methanol masers are pumped by infrared radiation and located close to high-mass young stellar objects \citep[within $1^{\prime\prime}$,][]{Cas2010}.

 The 6.7 GHz transition is an extensively studied Class \uppercase\expandafter{\romannumeral2} methanol maser and is one of the strongest masers known in our universe \citep[]{Men1988a,Men1988b,Nor1988,Men1991}. Unlike other maser species (OH and H$_2$O masers associated with both star forming regions and evolved stars), the 6.7 GHz methanol maser is only associated with massive star forming regions \citep[]{Min2003,Xu2008}. A large number of observations suggest that 6.7 GHz methanol maser is a powerful tool for studying massive star forming regions \citep[e.g.][]{Ell2006}.

 In the past few decades, there were several 6.7 GHz methanol maser surveys in our Galaxy, including targeted surveys \citep[e.g.][]{Men1991,Mac1992,Cas1995,Cas1996,Ell1996,Wal1996,Wal1997,Ell2007} and unbiased surveys \citep[e.g.][]{Gre2009,Gre2010,Gre2012}. Targeted surveys mainly observed infrared sources or star forming regions associated with known tracers such as H$_2$O or OH masers \citep[e.g.][]{Men1991}, while unbiased surveys are performed toward specific areas. The Methanol Multibeam (MMB) Survey conducted with the Parkes telescope is an unbiased survey of 6.7 GHz methanol maser in a relatively wide region of the Galactic plane \citep[186$^\circ$ $\le$ l $\le$ 60$^\circ$, $|b| \le$ 2$^\circ$,][]{Gre2009}. A catalog of about 519 6.7 GHz methanol masers is compiled by \citet{Pes2005}, and the MMB survey has detected 954 sources \citep[]{Cas2010,Gre2010,Cas2011,Gre2012,Bre2015}. Up to now, about 1000 Class
 \uppercase\expandafter{\romannumeral2} methanol maser sources have been detected in our Galaxy.

 Since star forming regions are considered to be located mainly in the Galactic plane, those 6.7 GHz methanol maser surveys in our Galaxy were conducted in the regions at low Galactic latitudes ($|b|<$ 2$^\circ$). In the catalog of \citet{Pes2005}, only 14 sources are at high Galactic latitudes, suggesting the decrease of methanol maser sources toward high latitude. The MMB survey was also conducted toward the Galactic plane with $|b| \le$ 2$^\circ$. Further systematic survey toward high-latitude regions are necessary to further check whether or not methanol masers are rare at high latitudes.

 Many previous interferometric observations (including connected element interferometers and VLBI) have revealed that some Class II methanol maser spots are distributed predominately in lines or arcs with a near-monotonic velocity gradient along the structure \citep[]{Nor1998,Phi1998,Bar2009,Bar2014,Bar2016,Fuj2014,Sug2016}. \citet{Bar2009} reported that about one-third of their sample sources show a new class of ``ring-like" methanol maser spot distribution. The new morphology suggests that methanol maser emission is produced by shocked material associated with a disk or a torus. However, some other observations also show that the methanol maser spot distribution has other morphologies, such as complex and pair morphologies \citep[]{Bar2014,Bar2016}. The complex methanol morphologies may be formed under the complicated cluster or neighbor environment around the massive star forming regions where the methanol masers locate. Therefore, a relatively clean environment with less clustering is helpful to better investigate the origin of 6.7 GHz methanol masers. The 6.7 GHz methanol maser sources at high-latitude ($|b|>$ 2$^\circ$) regions would provide potential candidates for such study, because they have quite a simple cluster environment compared to star forming regions at the Galactic plane.

 Several infrared surveys toward specific regions have been carried out, such as the Galactic Legacy Infrared Mid-Plane Survey Extraordinaire (GLIMPSE) infrared survey \citep[]{Ben2003} of the Galactic plane (-65$^\circ$ $\le$ l $\le$ 65$^\circ$, $|b|<$ 2$^\circ$) by \emph{Spitzer} and the survey carried out by the Midcourse Space Experiment (MSX) satellite ($|b|<$ 5$^\circ$). The all-sky \emph{Wide-Field Infrared Survey Explorer} (\emph{WISE}) catalogue covers the entire sky \citep[]{Wri2010}, and it is an up-to-date database. It observes at four mid-infrared bands: 3.4 $\mu$m, 4.6 $\mu$m, 12 $\mu$m and 22 $\mu$m, with resolutions of $6.1^{\prime\prime}$, $6.4^{\prime\prime}$, $6.5^{\prime\prime}$, $12^{\prime\prime}$, respectively. These all make \emph{WISE} a good database for the further methanol maser survey. In this paper, we report a survey of 6.7 GHz methanol masers toward a sample of \emph{WISE}-selected sources at high-latitude ($|b|>$ 2$^\circ$) regions. Sections 2 and 3, describe the sample selection and observations. We present the results in Section 4, followed by discussions in Section 5 and a summary in Section 6.

\section{Sample Selection}

We used the MMB survey catalog as a starting sample to select the \emph{WISE} point source candidates, which harbor the 6.7 GHz methanol maser. The MMB survey catalog has a position accuracy of $0.1^{\prime\prime}$ by ATCA \citep[]{Gre2009} and the astrometric precision of the \emph{WISE} point sources is better than $0.15^{\prime\prime}$ for high signal-to-noise ratio (S/N) sources and there is an additional error of about FWHM/(2 S/N) for S/N $<$ 20 sources \citep[]{Wri2010}, the FWHM represents the full width at half maximum of the point$-$spread function. The positions of the maser sources in the MMB catalog were determined at an accuracy of sub-arcsecond. The MMB catalog includes a total of 684 masers in the range of longitude $186^\circ$ $\leq$ $l$ $\leq$ $20^\circ$ and latitude $|b|<2^\circ$. We cross-matched the MMB catalog sources with the AllWISE catalog sources using $7^{\prime\prime}$ radius as the criteria. The AllWISE source catalog has accurate positions and four-band fluxes for $\sim$750 million sources extracted from its all-sky surveys \citep[]{Mai2011}. We select $7^{\prime\prime}$ radius as a matching criterion based on the spatial resolution of \emph{WISE} at shorter wavelengths and the possible spatial distribution of 6.7 GHz methanol masers \citep[usually less than 1$''$; e.g.][]{Cas2009}. As a result, we found that a total of 502 methanol maser sources in the MMB catalog have \emph{WISE} counterparts. There are five maser sources, each of which is matched with two \emph{WISE} point sources. For them, the angular separation of the two \emph{WISE} point sources associated with each of them is usually larger than $3^{\prime\prime}$ and we kept the closest \emph{WISE} point source only. Furthermore, we only considered those \emph{WISE} point sources (473 in total) with all the four-band flux measurements available.

 On the basis of the magnitude and color-color analysis for the selected 473 \emph{WISE} sources with 6.7 GHz MMB methanol maser associations, we further try to build up the criteria for the candidate \emph{WISE} sources that may have the 6.7 GHz methanol masers on the full sky. Fig. 1 shows the diagrams of the magnitudes of the \emph{WISE} four bands (3.4, 4.6, 12, and 22 $\mu$m) and the [3.4]--[4.6] versus [12]--[22] color of the selected sources. As a comparison, the corresponding distributions of those within a 1$^\circ$ diameter of the two locations of $l=90^\circ$, $b=70^\circ$ (denoted G90+70), and $l=0^\circ$, $b=0^\circ$ (denoted G00+00) are overlaid on Fig. 1. The comparison fields contain 13,627 and 18,939 \emph{WISE} sources for G90+70 and G00+00, respectively. Firstly, we select the \emph{WISE} sources with the magnitude criteria: $[3.4] < 14$mag; $[4.6] < 12$mag; $[12] < 11$mag; $[22] < 5.5$mag (see Fig. 1 (a) and Fig. 1 (b)). Almost all of (455/473=96\%) the known methanol masers lie toward the regions of these magnitude criteria; in contrast, for the comparison (G90+70) field, only about 0.4\% (53/13,627) of sources with the 3.4 and 4.5 $\mu$m data meet these criteria, and only one source satisfies the 12 and 22 $\mu$m criteria. Second, we further select the sample with the \emph{WISE} color criteria: $[3.4]-[4.6]>2$, and $[12]-[22]>2$. Toward this defined color region, we found that the majority of the methanol masers (330/455 = 73\%) fall in, whereas only nine \emph{WISE} sources from the comparison field of G90+70 locate (Fig. 1 (c)). Fig. 2 shows the histogram of the distribution of the angular separations between the 330 \emph{WISE} sources and their associated MMB masers. We found that most of them (288/330 = 87\%) are associated with the MMB masers within an angular scale of less than $3^{\prime\prime}$. Even for the comparison field of G00+00, where the background of the Galactic center is very complicated, there is still only a very small fraction (111/18,939=0.6\%) of the \emph{WISE} sources in this color region (Fig. 1 (d)). By applying the above criteria to the full sky, we found a total of $\sim$13,000 candidate \emph{WISE} sources. This is about 0.0017\% (13,000/750 $\times$ $10^6$) of the ALLWISE sources. Notably, in order to detect as many methanol masers as possible, we select ALLWISE sources satisfying the above criteria without considering the photometric accuracy factor of the \emph{WISE} sources. We also checked the \emph{WISE} sources with low S/Ns, e.g. S/N $<$ 3, associated with the MMB masers, and found that most of these sources also meet our selection criteria.

From these candidate \emph{WISE} sources, we further select a total of 3,348 \emph{WISE} sources as our targeted sample of the 6.7 GHz class \uppercase\expandafter{\romannumeral2} methanol masers to be observed by the Shanghai TianMa radio telescope (TMRT). This sample excluded the sources in the Parkes MMB survey and those with a declination of less than $-30^\circ$. In this targeted sample, 1473 sources locate at the high Galactic latitude region with $|b|>2^\circ$ (Table 1).

\section{Observations}
  The observations were performed between 2015 September and 2016 July with the TMRT in Shanghai, China. This telescope is a new fully steerable radio telescope with a diameter of 65m. We used a cryogenically cooled receiver covering the frequency range of 4$-$8 GHz and the Digital Backend System (DIBAS) to receive and record signals. The DIBAS is an FPGA-based spectrometer that is designed on the base of the Versatile GBT Astronomical Spectrometer \citep[VEGAS;][]{Bus2012}. The rest frequency of the methanol 5$_1$ $\rightarrow$ 6$_0$ A$^+$ transition line (6.6685192 GHz) was covered with a spectral window of a bandwidth of 23.4 MHz. The window has 16,384 channels and the spectral resolution is 1.431 kHz, corresponding to a velocity resolution of $\sim$ 0.09 km s$^{-1}$. The temperature of this system is about 20$-$30 K. The aperture efficiency of the TMRT is $\sim$ 55\% and the corresponding sensitivity is 1.5 Jy K$^{-1}$. The beam size is $\sim$ $3^{\prime}$ (HPBW) at the frequency of 6.7 GHz. The flux densities for the sources may have an uncertainty of less than 20\%, estimated from the observed variation of the calibrators.

  A position-switching mode was used in our observations. Each source was observed with two ON$-$OFF cycles. Procedures of both the ON position and the OFF position in each cycle take $\sim$ 2 minutes. For each source, the OFF position was set to (0.0$^\circ$, $-$0.4$^\circ$) from the ON position in (R.A., decl.).

  We used the GILDAS/CLASS package to process the observed spectral data. We fitted and subtracted the linear baseline of the spectrum. The root-mean-square (rms) noise level is about 20$-$40 mJy.

  For the detected methanol maser sources (12 in total, see below), we used the on-the-fly (OTF) mode \citep{Don2016} to map a large region (usually 5$^{\prime}$) to obtain a position of the methanol maser. The TMRT under the OTF mode moved in a given region along the direction of R.A. first, then the direction of decl. Only one spectral window is used to cover the methanol maser.

\section{Results}
\subsection{Detections}

We detected 6.7 GHz methanol masers toward positions of 17 \emph{WISE} point sources, within actually 12 sources. The properties of these detected methanol maser sources are listed in Table 2. Three sources are newly detected and Fig. 3 shows their spectra. We confirmed that another nine methanol maser sources detected by the TMRT observations were already reported after taking their similar positions (within 1$^{\prime}$) and spectral velocity ranges into account. Their spectra are shown in Fig. 4. Below are short descriptions of these 12 sources.

\subsubsection{Three newly detected sources}

 \textbf{\emph{G97.527+3.184}}. This source has three peaks at $-$73.0, $-$70.9 and $-$69.7 km s$^{-1}$, within a velocity range of $-$78.4 km s$^{-1}$ to $-$67.6 km s$^{-1}$.

 \textbf{\emph{G110.196+2.476}}. This weak source has a peak flux density of 0.5 Jy at $-$56.9 km s$^{-1}$.

 \textbf{\emph{G137.068+3.002}}. This source has a peak flux density of 1.6 Jy at $-$59.0 km s$^{-1}$.

 Compared to previously known sources, all the three new sources show relatively weak maser intensity. G110.196+2.476 and G137.068+3.002 have a simple spectral feature.

 G97.527+3.184 has also been studied by \citet[]{Hac2015} through the 22 GHz H$_2$O maser and it is located in the Outer Arm. So, this source may provide a good candidate to study the massive star formation that occurred in the regions at the edge of our Galaxy.

Using the multi-band infrared data, including the \emph{WISE} and \emph{Herschel},
we can estimate the related physical parameters for the newly
detected sources. However, in the three sources, only one source,
G110.196+2.476, is covered by \emph{Herschel}.

We estimated the dust temperature of G110.196+2.476 from its
spectral energy distributions (SED) in the \emph{WISE} 22 $\mu$m, \emph{Herschel}
PACS 70 and 160 $\mu$m, and SPIRE 250, 350, and 500 $\mu$m. The SED
is fitted using a gray-body emission model \citep{hildebrand83}
\begin{equation}
S_{\nu}  = \kappa_\nu B_{\nu}(T_{\rm d}) \Omega \mu m_{\rm H} N_{\rm tot}/g \\
        = \frac{\kappa_{\nu} B_{\nu}(T_{\rm d}) M_{\rm core} }{g D^2},
\end{equation}
wherein $S_{\nu}$ is the flux density at the frequency $\nu$.
$\Omega$ is the solid angle of the core or the selected area.
$B_{\nu}(T_{\rm d})$ is the Planck function of the dust temperature
$T_{\rm d}$. $\mu=2.33$ is the mean molecular weight
\citep{myers83}, $m_{\rm H}$ is the mass of the hydrogen atom,
$g=100$ is the gas-to-dust mass ratio, $N_{\rm tot}$ is the column density of the core, $D=5.1$ kpc is the source
distance. $\kappa_\nu$ is the dust opacity and is expected to vary
with the frequency in the form $\kappa_\nu=\kappa_{\rm 230
GHz}(\nu/{\rm 230 GHz})^{\beta}$, with the reference value of
$\kappa_{\rm230 GHz}=0.9$ cm$^2$ g$^{-1}$, as adopted from dust
model for the grains with coagulation for $10^5$ years with accreted
ice mantles at a density of $10^6$ \citep{ossenkopf94}.

The continuum emission of the core has a compact and isolated shape
throughout all the wavebands. The flux density can be reproduced by a
single temperature component, as shown in Fig. 5. The derived
physical parameters are $M_{\rm core}=14~M_\odot$, $T_{\rm d}=49$ K
and $\beta=1.4$ for G110.196+2.476. Therefore, the new detection of
the 6.7 GHz methanol maser from this source is still associated with
the massive star-formation process.

\subsubsection{9 previously detected sources}

 \textbf{\emph{G16.872$-$2.154, G16.883$-$2.186}}. The methanol maser emission detected from these two positions actually should be from a common position close to G16.872$-$2.154. Previous observations from \citet{Cas1995}, \citet{Wal1997,Wal1998}, and \citet{Szy2000} also detected the 6.7 GHz maser  emission from G16.872$-$2.154. Comparing our spectra to those in \citet{Szy2000}, we found that these masers have similar velocity features, but peak flux density changed from 23 Jy to 16.0 Jy in G16.872$-$2.154.

 \textbf{\emph{G17.021$-$2.402}}. This source was detected by \citet{Szy2000}, named 18277$-$1517. Both observations detected maser emission at the same velocity range, but flux density changed significantly. The peak flux density was 3.7 Jy at 23.3 km s$^{-1}$ and 8.9 Jy at 20.6 km s$^{-1}$ in \citet{Szy2000} and our observations, respectively.

 \textbf{\emph{G78.122+3.633}}. This source was detected by \citet{Mac1992}, \citet{Sly1999}, \citet{Szy2000}, and \citet{Min2000,Min2001}, named 20126+4104. TMRT measured a peak flux density of 39.9 Jy at $-$7.7 km s$^{-1}$, similar to 38 Jy at $-$6.1 km s$^{-1}$ in \citet{Szy2000}.

 \textbf{\emph{G108.184+5.518}}. It was detected by \citet{Mac1998}, \citet{Sly1999}, and \citet{Szy2000}. This source has a peak flux density of 49.2 Jy at $-$10.7 km s$^{-1}$ by the TMRT. In \citet{Szy2000}, it (named 22272+6358) had a peak flux density of 69.4 Jy at $-$11.1 km s$^{-1}$.

 \textbf{\emph{G109.839+2.134, G109.868+2.119}}. The spectra of these two separate sources show that most maser emission is from G109.868+2.119, which has a very strong peak flux density of 766.7 Jy at $-$2.4 km s$^{-1}$. G109.868+2.119 was detected by \citet{Sly1999} and \citet{Szy2000}. \citet{Szy2000} reported G109.868+2.119 (named 22543+6145) with a feature of peak flux density of 815 Jy.

 \textbf{\emph{G123.035$-$6.355, G123.050$-$6.310}}. The maser emission detected toward these two positions are from the same source: G123.050$-$6.310. \citet{Sly1999} and \citet{Szy2000} detected G123.050$-$6.310. \citet{Szy2000} detection, with the name 00494+5617, has a similar spectrum but with a peak flux density of 24 Jy, compared with our detection of 66.9 Jy.

 \textbf{\emph{G173.482+2.446}}. This 6.7 GHz methanol maser site was detected by \citet{Men1991}, \citet{Szy2000} and \citet{Min2000}, named S231. In \citet{Szy2000}, it showed a much stronger flux density (a peak flux density of 208 Jy) than our detection (41.5 Jy), with a quite different spectrum.

 \textbf{\emph{G173.596+2.823, G173.617+2.883}}. The spectra at both positions clearly show that most emission is from G173.617+2.883, with a velocity range of $-$24.8 km s$^{-1}$ to $-$18.1 km s$^{-1}$. G173.617+2.883 was detected by \citet{Szy2000}, and was named 05382+3527.

 \textbf{\emph{G213.706$-$12.602, G213.752$-$12.615}}. These two positions have features at the same velocity range and the emission is mostly from G213.706$-$12.602. G213.706$-$12.602 was detected by \citet{Szy2000}, named 06053$-$0622, with a peak flux density of 166 Jy at 10.5 km s$^{-1}$. However, we found two peaks with peak flux densities of $\sim$120 Jy at both 10.5 and 12.7 km s$^{-1}$. This source, named Mon R2, was also detected in \citet{Men1991}, \citet{Cas1995}, \citet{Wal1997,Wal1998}, and \citet{Min2001}.

\subsection{Maser positions determined from the OTF observations}
 The center coordinates and the side length of the square regions chosen for the OTF observations toward these 12 sources are listed in Table 3. Fig. 6 shows the velocity-integrated intensity map from the OTF observation and the 6.7 GHz methanol maser spectra in the fitting center toward G97.527+3.184, the peak positions in each region can be found in Table 3. Considering the pointing error of the telescope (about $10^{\prime\prime}$) and the fitting error from the OTF mapping observation (typically $10^{\prime\prime}$), we can obtain a positional accuracy at a level of better than $20^{\prime\prime}$ with the TMRT OTF observations.

 \citet{Min2000,Min2001} and \citet{Hu2016} have studied eight of these sources through interferometric observations with the European VLBI Network (EVN), the Very Long Baseline Array, and the Very Large Array (VLA). For comparison, the interferometric positions of these eight sources are also listed in Table 3, along with their position difference from our OTF measurements. The positions of the six sources (G17.021$-$2.402, G78.122+3.633, G108.184+5.518, G109.868+2.119, G123.050$-$6.310 and G173.482+2.446) determined from our OTF observations have offsets in the range of 10$^{\prime\prime}$ $\sim$ 20$^{\prime\prime}$, with regard to the positions determined from the interferometric observations. This is consistent with a positional uncertainty of less than 20$^{\prime\prime}$ from the OTF observations. For G16.872$-$2.154 and G213.706$-$12.602, the position accuracy determined from the OTF observations are 3.0$^{\prime\prime}$ and 10$^{\prime\prime}$, respectively.

\section{Discussions}
\subsection{Rarity of high-latitude methanol masers}
In total, 17 methanol maser sources at high latitudes, including known sources from the \citet{Pes2005} catalog and those first reported in this work, have been detected to date. The distribution of 6.7 GHz methanol maser sources at high latitude is shown in Fig. 7. It can be seen that about three-fourths (13 of 17 sources) are located at Galactic latitudes of 2$^\circ\sim4^\circ$ and $-2^\circ\sim-4^\circ$. According to statistics \citep[]{Pes2005,Cas2010,Gre2010,Cas2011,Gre2012,Bre2015}, there are about 1000 Class
 \uppercase\expandafter{\romannumeral2} methanol maser sources detected in our Galaxy, suggesting a $\sim1.7\%$ fractional distribution of high-latitude sources.

The latitude distribution of 519 methanol masers listed in Figure 3 in \citet{Pes2005} has a Gaussian fit with an FWHM of 0.52$^\circ$, consistent with the rarity ($\sim1.7\%$) of 6.7 GHz methanol masers at high latitude. However, this situation may be related to the fact that previous methanol maser surveys were mainly concentrated near the Galactic plane. Our survey detected twelve 6.7 GHz maser sources toward the region of 0$^\circ$ $\le$ l $\le$ 220$^\circ$ and $|b| > 2^\circ$. Assuming a uniform distribution of methanol masers along the galactic longitude, then a total number of 20 is estimated to be detected in the whole high-latitude region (0$^\circ$ $\le$ l $\le$ 360$^\circ$, $|b| > 2^\circ$) toward the \emph{WISE}-selected sample. Because the detection rate is estimated to $\sim 73\%$ (see $\S$ 2), the expected number of methanol masers would be 27 in the region of $|b| > 2^\circ$. Thus, our survey strongly suggests that the 6.7 GHz methanol maser is really rare in the high-latitude region.

\subsection{Profile variability of methanol masers}
 As described in Section 4.1.2, 9 of 12 TMRT detected methanol maser sources were previously observed by \citet{Szy2000} with the 32m Toru\'n radio telescope with an HPBW of $5.5^{\prime}$ at 6.7 GHz. These sources are not accurately pointed from the same coordinate position, which may cause differencec in flux density. Our discussion will mainly focus on the variations of their spectral profiles. We also list the source names in parenthesis from \citet{Szy2000} at the beginning of the description of each source.

 \textbf{\emph{G16.872$-$2.154 (18265$-$1517)}}. The spectra of both G16.872$-$2.154 and 18265$-$1517 show structures with several features and, the peak flux densities are 19.3 Jy and 23 Jy, respectively. The features at about 15.0, 17.3, and 18.6 km s$^{-1}$ of G16.872$-$2.154 are quite similar to those at about 14.7, 17.0, and 18.3 km s$^{-1}$ of 18265$-$1517 with a velocity offset of 0.3 km s$^{-1}$. Such an offset may not be significant compared to the reported error of $\pm$0.4 km s$^{-1}$ in the absolute radial velocity.

 \textbf{\emph{G17.021$-$2.402 (18277$-$1517)}}. From the spectra of this source and 18277$-$1517, their peak flux densities are 9.2 Jy and 3.7 Jy, respectively. Profiles have changed a lot. In the \citet{Szy2000} observation, there was a feature at about 23.3 km s$^{-1}$ and an obscure feature at about 20.6 km s$^{-1}$. However, our spectra show features at 19.4, 20.6, 21.7, 23.3, and 24.4 km s$^{-1}$.

 \textbf{\emph{G78.122+3.633 (20126+4104)}}. The peak flux densities of G78.122+3.633 and 20126+4104 are 40.9 Jy and 38 Jy, respectively. Except for a stable feature at $-$6.1 km s$^{-1}$, 20126+4104 did not show any other features seen in G78.122+3.633 at $-$8.2, $-$7.6, $-$6.5, and $-$5.0 km s$^{-1}$.

 \textbf{\emph{G108.184+5.518 (22272+6358)}}. The peak flux densities of G108.184+5.518 and 22272+6358 are 49.2 and 91 Jy, respectively. Compared to \citet{Szy2000}, features at about $-$10.7 km s$^{-1}$ and $-$10 km s$^{-1}$ remain unchanged in our observation, whereas a new feature at about $-$12.7 km s$^{-1}$ arose in our observation.

 \textbf{\emph{G109.868+2.119 (22543+6145)}}. We can see that the spectra of G109.868+2.119 and 22543+6145 have similar features and flux densities. It seems that this source has not changed a lot.

 \textbf{\emph{G123.050$-$6.310 (00494+5617)}}. From the spectra of G123.050$-$6.310 and 00494+5617, we can see that the peak flux densities are 97.2 Jy and 24 Jy, respectively. The features at $-$36.0 km s$^{-1}$ and $-$30.8 km s$^{-1}$ have similar flux density ratios, but become stronger now.

 \textbf{\emph{G173.482+2.446 (05358+3543)}}. The peak flux densities of G173.482+2.446 and 05358+3543 are 41.5 and 256 Jy, respectively. We can clearly see features at $-$14.7, $-$13, and $-$10.5 km s$^{-1}$ in both observations. Moreover, there is a new feature at about $-$7.5 km s$^{-1}$. In addition, the peak flux densities have decreased a lot.

 \textbf{\emph{G173.617+2.883 (05382+3547)}}. The peak flux density has changed from 7.5 Jy in 05382+3547 to 4.5 Jy in G173.617+2.883 and a new feature arose at $-$20.1 km s$^{-1}$.

 \textbf{\emph{G213.706$-$12.602 (06053$-$0622)}}. The peak flux densities of G213.706$-$12.602 and 06053$-$0622 are 127 Jy and 166 Jy, respectively. The features at about 11.6 km s$^{-1}$ and 12.5 km s$^{-1}$ disappeared and the new feature at about 12.7 km s$^{-1}$ showed up with a flux density as strong as that of the feature at about 10.5 km s$^{-1}$.

The 3$\sigma$ noise level was typically 1.5$-$1.9 Jy in \citet{Szy2000}, and ours is less than 0.2Jy which is much better.  This may explain the appearance of some new features in the TMRT observations.

 Compared with \citet{Szy2000} observations, except for G109.868+2.119, 8 of 9 sources have shown significant profile changes. Many 6.7 GHz methanol maser sources have been observed to be variable \citep[e.g.][]{Goe2004,Goe2009,Goe2014,Wal2009,Wal2016,Szy2011,Szy2015,Mas2015,Mas2016}. Several possibilities might cause such variabilities: (i) pulsation of a young high-mass star \citep[]{Ina2013,San2015}, (ii) a colliding-wind binary system \citep[]{Wal2009}, and (iii) rotating spiral shocks in the gap region of a binary system \citep[]{Par2014}.

\subsection{WISE infrared environment source associations}

The infrared three-color images of these 12 TMRT detected methanol maser sources are shown in Fig. 8 with the fitting centers of methanol emission determined from the TMRT OTF observations marked. Two sources (G17.021$-$2.402 and G173.617+2.883) do not have bright counterparts in the \emph{WISE} infrared image. The positions of three sources (G108.184+5.518, G110.196+2.476, and G173.482+2.446), determined from the TMRT OTF observations are within $20^{\prime\prime}$ from the nearby bright infrared sources. All of the remaining seven methanol maser sources locate in the bright \emph{WISE} sources. Considering the $20^{\prime\prime}$ positional accuracy of TMRT OTF observations (see $\S$ 4.2), 10 of the 12 detected sources are associated with bright \emph{WISE} sources.

 Furthermore, with more accurate interferometric positions for eight sources (see $\S$ 4.2), we found that seven sources are well associated with the \emph{WISE} sources, with an exception of G17.021$-$2.402. For six out of the eight sources (the two exceptions are G16.872$-$2.154 and G123.050$-$6.310), we also found that methanol masers are located very close to the \emph{WISE} sources, which were targeted in our survey. Therefore, we could also refine a methanol maser source at a positional accuracy of about $10^{\prime\prime}\sim20^{\prime\prime}$, by solely matching to the \emph{WISE} source positions. This is especially important for those maser sources without any high accuracy position measurements, such as those from surveys toward a sample selected from \emph{WISE} sources at a low-latitude region.

 \subsection{Spatial distribution of methanol maser spots}

 The 6.7 GHz methanol maser distributions toward eight sources have been studied with interferometric observations (see $\S$ 4.2) and we can compare the distribution of the maser spots from all these eight sources obtained from the JVLA at C-array configuration (Hu et al. 2016). There are four sources, G16.872$-$2.154, G17.021$-$2.402, G78.122+3.633, and G213.706$-$12.602, showing the maser distribution in a linear morphology, and three sources, G108.184+5.518, G109.868+2.119, and G123.050$-$6.310 seem to distribute along a ring-like structure. In addition, the distribution of G173.482+2.446 looks complex. The linear or ring$-$shaped distribution of the masers may be associated with jet/outflow or disk/torus structures \citep[e.g.][]{Bar2016} from an isolated massive young stellar objects. However, the complex structure of the masers should be affected by the neighborhoods in a complicated cluster environment in the massive star forming regions.

 Of the 63 low-latitude sources with the EVN imaging of 6.7 GHz methanol masers \citep{Bar2009,Bar2014,Bar2016}, 1 source shows simple morphology, 11 sources show ring morphology, 13 sources show linear morphology, 5 sources show arched morphology, 29 sources show complex morphology, and 4 sources show pair morphology. This proportion is quite different from our high-latitude ones, especially for the complex morphology group: only one in 8 sources from our sample, compared to about half (29/63) in Bartkiewicz et al. (2009, 2014, 2016) samples. This is likely to be due to the fact that the sources at the low latitudes usually have more complicated cluster environments. The physical and kinematic contributions from the neighborhoods around the host of methanol masers would complicate the morphology of methanol masers. This makes it difficult to use a methanol maser as a useful probe to the physics and kinematics that are solely associated with its host. From this point of view, the detected methanol maser sources at high latitudes from our observations would serve as better candidates for studying the physics and kinematics during the massive star formation in our further work using VLA observations.

\section{Summary}
 Using the newly built TMRT in Shanghai, we undertook a systematic survey of 6.7 GHz methanol masers toward high-latitude ($|b|>$ 2$^\circ$) targets in our Galaxy. The target sample contains 1473 high-latitude sources selected from the \emph{WISE} source catalog. We detected 12 methanol masers with 3 new detections, suggesting that the 6.7 GHz methanol masers are really rare at the high-latitude region. The peak flux densities of the detected sources are from 0.9 to 838.8 Jy, while new detections have peak flux densities in the range of 0.9 $\sim$ 1.8 Jy. Comparing with previous detections toward nine sources, we found that the spectra and flux densities of eight sources have changed significantly. Combining our OTF observations and previous interferometric observations, we concluded that most methanol masers are associated with bright \emph{WISE} point sources in the infrared images. The published interferometric observations revealed that almost all of these high-latitude sources do not have complex morphology in the methanol maser spot distribution, in contrast to those at low latitudes. Therefore, high-latitude methanol masers might serve as an efficient tracer for studying physics and kinematics of their hosts during the massive star formation.

\section*{Acknowledgements}

 We are thankful for the assistance from the operators of the TMRT during the observations and for helpful comments that improved the manuscript. This work was supported by the National Natural Science Foundation of China (11590780, 11590781, and 11590784), the Strategic Priority Research Program ``The Emergence of Cosmological Structures" of the Chinese Academy of Sciences (CAS), Grant No. XDB09000000, the Knowledge Innovation Program of the Chinese Academy of Sciences (Grant No. KJCX1-YW-18), the Scientific Program of Shanghai Municipality (08DZ1160100), and Key Laboratory for Radio Astronomy, CAS.

\newpage
\begin{sidewaystable}[h]
\centering
{\scriptsize
\caption{Selected Source Catalog}
\centerline{\begin{tabular}{c c c c ccccc c c}
\hline
 Number & l & b & R.A. & Decl. & w1 (3.4 $\mu m$) & w2 (4.6 $\mu m$) & w3 (12 $\mu m$) & w4 (22 $\mu m$) & w1$-$w2 & w3$-$w4 \\
 & ($^{\circ}$) & ($^{\circ}$) & (J2000) & (J2000) & (mag) & (mag) & (mag) & (mag) & (mag) & (mag) \\
 &  &  & (h m s) & ($^{\circ}$ $^{\prime}$ $^{\prime\prime}$) &  &  &  &  &  & \\
 (1) & (2) & (3) & (4) & (5) & (6) & (7) & (8) & (9) & (10) & (11) \\
\hline
 1 & 0.125 & 5.111 & 17:26:29.0 & $-$26:04:57.1 & 8.75 & 6.24 & 2.22 & $-$0.08 & 2.51 & 2.30 \\
 2 & 0.291 & 3.992 & 17:31:03.7 & $-$26:33:33.5 & 13.04 & 10.45 & 6.21 & 2.70 & 2.59 & 3.51 \\
 3 & 0.472 & $-$2.192 & 17:55:21.8 & $-$29:39:12.9 & 11.43 & 7.24 & 3.10 & 0.52 & 4.18 & 2.58 \\
 4 & 0.538 & $-$2.720 & 17:57:37.2 & $-$29:51:38.7 & 8.19 & 5.84 & 5.84 & 2.88 & 2.34 & 2.96 \\
 5 & 0.689 & 2.141 & 17:39:01.0 & $-$27:13:23.7 & 8.95 & 5.98 & 2.66 & 0.45 & 2.96 & 2.21 \\
 6 & 0.819 & 2.517 & 17:37:53.9 & $-$26:54:44.8 & 7.59 & 5.23 & 3.29 & 1.04 & 2.36 & 2.24 \\
 7 & 0.906 & $-$2.147 & 17:56:11.2 & $-$29:15:20.6 & 6.94 & 4.61 & 2.72 & 0.68 & 2.34 & 2.04 \\
 8 & 0.984 & 6.051 & 17:25:06.8 & $-$24:51:00.3 & 9.56 & 6.34 & 3.49 & 1.05 & 3.21 & 2.44 \\
 9 & 1.101 & $-$3.114 & 18:00:29.0 & $-$29:34:08.0 & 8.46 & 5.90 & 2.85 & 0.73 & 2.56 & 2.12 \\
 10 & 1.227 & 2.005 & 17:40:48.8 & $-$26:50:19.2 & 9.61 & 5.92 & 2.87 & 0.39 & 3.70 & 2.48 \\
 &  &  &  &  &       ......        &  &  &  &  & \\
\hline
\end{tabular}}}
\tablecomments{
  Column 1: source number; Columns 2 - 5: the positions (gl, gb, R.A., decl.); Columns 6 - 9: magnitudes in the 4 \emph{WISE} bands; Columns 10 - 11: the \emph{WISE} color of the sources. (This table is available in its entirety in machine-readable form.)}
\end{sidewaystable}

\newpage
\begin{sidewaystable}[h]
\centering
{\scriptsize
\caption{Properties of the TMRT detected methanol masers toward \emph{WISE} point sources.}
\centerline{\begin{tabular}{ccccccccccc}
\hline
Name & R.A. & Decl. & Epoch & $\Delta V$ & $v_p$ & $S_p$ & $S_i$ & Distance & Ref. & Other Names\\
 (l, b)  & (J2000) & (J2000) & yy/mm/dd & (km s$^{-1}$) & (km s$^{-1}$) & (Jy) & (Jy km s$^{-1}$) & (kpc) &  &  \\
 ($^{\circ}$, $^{\circ}$)  & (h m s) & ($^{\circ}$ $^{\prime}$ $^{\prime\prime}$) &  &  &  &  &  &  &  &  \\
 (1) & (2) & (3) & (4) & (5) & (6) & (7) & (8) & (9) & (10) & (11) \\
\hline
 G16.872$-$2.154 & 18:29:24.3 & $-$15:15:30.0 & 16/01/17 & 14.4, 20.0 & 15.1 & 15.8 & 4.7 & 1.2$^{a}$ & C,W,S & L379$-$IRS3 \\
       &      &     &  &     & 17.4 & 16.0 & 4.9 &     &       & \\
 G16.883$-$2.186 & 18:29:32.6 & $-$15:15:46.4 & 16/01/17 & 14.7, 18.1 & 15.0 & 1.1 & 0.3 & & C,W,S & \\
       &      &   &   &     & 17.4 & 1.2 & 0.3 &    &    &  \\
 &  &  &  &  &  &  &  &  & & \\
 G17.021$-$2.402 & 18:30:36.2 & $-$15:14:26.9 & 16/06/15 & 18.8, 24.7 & 20.6 & 8.9 & 6.1 & 1.8$^{a}$ & S & L379$-$IRS2\\
 &  &  &  &  &  &  &  &  & & \\
 G78.122+3.633 & 20:14:26.2 & +41:13:32.1 & 16/06/09 & $-4.6, -8.8$ & $-$7.7 & 39.9 & 63.1 & 2.3$^{a}$ & G,V,S,M,M1 & IRAS 20126+4104 \\
 &  &  &  &  &  &  &  &  & & \\
 G97.527+3.184 & 21:32:11.3 & +55:53:39.6 & 16/01/11 & $-78.4, -67.6$ & $-$73.0 & 0.8 & 0.2 & 7.5$^{b}$ & N & IRAS 21306+5540 \\
       &      &      &    & & $-$70.9 & 1.0 & 0.4 &      &  &  \\
       &      &      &    & & $-$69.7 & 0.7 & 0.3 &      &  &  \\
        &  &  &  &  &  &  &  &  & & \\
 G108.184+5.518 & 22:28:51.5 & +64:13:40.2 & 15/12/12 & $-13.1, -9.2$ & $-$10.7 & 49.2 & 18.6 & 0.8$^{a}$ & G8,V,S & L1206 \\
 &  &  &  &  &  &  &  &  & & \\
 G109.839+2.134 & 22:55:58.8 & +62:02:02.5  & 16/01/18 & $-4.9, -1.3$ & $-$2.4 & 187.8 & 73.2 & & K,S,M1 & \\
 G109.868+2.119 & 22:56:15.8 & +62:01:59.4 & 16/01/18 & $-4.9, -1.4$ & $-$2.4 & 766.7 & 294.3 & 0.8$^{a}$ & K,S,M1 & Cep A \\
 &  &  &  &  &  &  &  &  & & \\
 G110.196+2.476 & 22:57:29.8 & +62:29:45.1 & 16/01/18 & $-57.2, -56.4$ & $-$56.9 & 0.5 & 0.2 & 5.1$^{c}$ & N & \\
 &  &  &  &  &  &  &  &  & & \\
 G123.035$-$6.355 & 00:52:11.0 & +56:30:58.8 & 16/03/06 & $-36.5, -27.6$ & $-$29.3 & 2.6 & 1.2 & & S & \\
 G123.050$-$6.310 & 00:52:17.2 & +56:33:42.6 & 16/03/06 & $-36.5, -27.5$ & $-$29.3 & 66.9 & 29.2 & 2.4$^{a}$ & S & NGC281/S184\\
 &  &  &  &  &  &  &  &  &  & \\
 G137.068+3.002 & 02:58:13.3 & +62:20:31.9 & 16/03/05 & $-59.8, -58.1$ & $-$59.0 & 1.6 & 1.2 & 3.0$^{c}$ & N & \\
 &  &  &  &  &  &  &  &  & & \\
 G173.482+2.446 & 05:39:13.0 & +35:45:50.9 & 16/02/05 & $-15.3, -7.2$ & $-$13.0 & 41.5 & 17.7 & 1.7$^{a}$ & K,S,M & S231/IRAS 05358+3543 \\
 &  &  &  &  &  &  &  &  & & \\
 G173.596+2.823 & 05:41:05.4 & +35:52:02.5 & 16/04/14 & $-24.8, -18.1$ & $-$24.2 & 0.6 & 0.3 &  & S & \\
 G173.617+2.883 & 05:41:24.1 & +35:52:50.6 & 16/04/14 & $-24.8, -18.2$ & $-$24.3 & 4.5 & 2.5 & 1.8$^{d}$ & S & S235\\
 &  &  &  &  &  &  &  &  &  & \\
 G213.706$-$12.602 & 06:07:46.8 & $-$06:23:08.3 & 16/03/07 & 9.7, 13.8 & 10.5 & 121.8 & 45.7 & 0.8$^{a}$ & K,C,W,S,M1 & Mon R2\\
        &      &      &   &  & 12.7 & 121.4 & 57.3 &         &  & \\
 G213.752$-$12.615 & 06:07:48.8 & $-$06:25:55.3 & 16/03/07 & 9.7, 13.7 & 10.5 & 4.4 & 1.6 &  & K,C,W,S,M1 & \\
        &      &      &     &  & 12.7 & 4.4 & 2.0 &         &  & \\
\hline
\end{tabular}}}
\tablecomments{
  Note. Column 1: source name; Columns 2 - 3: the targeted positions for the TMRT observations; Column 4: the epoch of observation; Column 5: the velocity interval of the maser emission; Column 6: the velocity of peak emission; Column 7: the peak flux density; Column 8: the integrated flux density; Column 9: the distance with references a \citep[]{Hu2016}, b \citep[]{Hac2015}, c \citep[estimated kinematic distances by using the Galactic rotation model of][]{Rom2009}, d \citep[]{Dew2016}; Column 10: the discovery references C \citep[]{Cas1995}, G \citep[]{Mac1992}, G8 \citep[]{Mac1998}, K \citep[]{Men1991}, N (newly discovered), M \citep[]{Min2000}, M1 \citep[]{Min2001}, S \citep[]{Szy2000}, V \citep[]{Sly1999}, and W \citep[]{Wal1997,Wal1998}; Column 11: other names of the source.}
\end{sidewaystable}

\newpage
\begin{sidewaystable}[h]
{\tiny
\caption{The OTF observations}
\centering
\centerline{\begin{tabular}{cccccccccccccccc}
\hline
 Name & \multicolumn {4} {c} {OTF Parameters} & & \multicolumn {5} {c} {Properties of the Fitting Centers} & &  \multicolumn {3} {c} {Properties of 8 Sources by Interferometric Observation} & Difference\\
  \cline{2-5} \cline{7-11} \cline{13-15}
 & R.A. & Decl. & Side & Epoch & & R.A. & Decl. & $v_p$ & $S_p$ & $S_i$ & & R.A. & Decl. & Ref. & \\
 (l, b) & (J2000) & (J2000) & Length & yy/mm/dd & & (J2000) & (J2000) & (km s$^{-1}$) & (Jy) & (Jy km s$^{-1}$) & & (J2000) & (J2000) & & ($^{\prime\prime}$) \\
 ($^{\circ}$, $^{\circ}$) & (h m s) & ($^{\circ}$ $^{\prime}$ $^{\prime\prime}$) & ($^{\prime}$) & & & & (h m s) & ($^{\circ}$ $^{\prime}$ $^{\prime\prime}$) & & & & (h m s) & ($^{\circ}$ $^{\prime}$ $^{\prime\prime}$) & & \\
 (1) & (2) & (3) & (4) & (5) & & (6) & (7) & (8) & (9) & (10) & & (11) & (12) & (13) & (14) \\
\hline
 G16.872$-$2.154 & 18:29:24.3 & $-$15:15:30.0 & 6 & 16/07/14 & & 18:29:24.20 & $-$15:16:03.6 & 15.1 & 19.0 & 5.3 & & 18:29:24.407 & $-$15:16:04.24 & H & 3.0 \\
 &  &  &  & &  &  &  & 17.4 & 19.3 & 5.8 &  &  & &  & \\
 G17.021$-$2.402 & 18:30:36.2 & $-$15:14:26.9 & 5 & 16/07/16 & & 18:30:36.96 & $-$15:14:41.4 & 20.6 & 9.2 & 6.3 & & 18:30:36.292 & $-$15:14:28.49 & H & 16.1 \\
 G78.122+3.633 & 20:14:26.2 & +41:13:32.1 & 5 & 16/07/14 & & 20:14:26.96 & +41:13:49.0 & $-$7.7 & 40.9 & 64.7 & & 20:14:26.047 & +41:13:32.58 & M,M1,H & 19.4\\
 G97.527+3.184 & 21:32:11.3 & +55:53:39.6 & 5 & 16/07/14 & & 21:32:11.18 & +55:53:32.0 & $-$73.0 & 0.9 & 0.3 &  & &  &  & \\
  &  &  &  &  & &  &  & $-$70.9 & 1.0 & 0.4 &  &  & &  & \\
  &  &  &  &  & &  &  & $-$69.7 & 0.8 & 0.3 &  &  & &  & \\
 G108.184+5.518 & 22:28:51.5 & +64:13:40.2 & 5 & 16/07/15 & & 22:28:52.07 & +64:13:58.3 & $-$10.7 & 50.0 & 19.8 & & 22:28:51.386 & +64:13:41.22 & H & 17.7 \\
 G109.868+2.119 & 22:56:15.8 & +62:01:59.4 & 6 & 16/07/15 & & 22:56:18.96 & +62:01:39.6 & $-$2.4 & 838.8 & 327.9 & & 22:56:18.095 & +62:01:49.45 & M1,H & 11.6 \\
 G110.196+2.476 & 22:57:29.8 & +62:29:45.1 & 5 & 16/07/16 & & 22:57:32.88 & +62:29:51.9 & $-$56.9 & 0.9 & 0.4 &  & &  &  & \\
 G123.050$-$6.310 & 00:52:17.2 & +56:33:42.6 & 6 & 16/07/16 & & 00:52:25.45 & +56:33:53.8 & $-$29.3 & 97.2 & 43.1 & & 00:52:24.200 & +56:33:42.98 & H & 15.0\\
 G137.068+3.002 & 02:58:13.3 & +62:20:31.9 & 5 & 16/07/18 & & 02:58:14.91 & +62:20:42.8 & $-$59.0 & 1.8 & 1.6 &  &  & &  & \\
 G173.482+2.446 & 05:39:13.0 & +35:45:50.9 & 5 & 16/07/18 & & 05:39:12.84 & +35:46:09.1 & $-$13.0 & 42.3 & 21.8 & & 05:39:13.059 & +35:45:51.29 & M,H & 18.0 \\
 G173.617+2.883 & 05:41:13.0 & +35:51:02.9 & 7 & 16/07/23 & & 05:41:22.75 & +35:52:31.5 & $-$18.2 & 5.0 & 3.1 & &  &  &  & \\
 G213.706$-$12.602 & 06:07:46.8 & $-$06:23:08.3 & 6 & 16/07/23 & & 06:07:48.41 & $-$06:22:51.3 & 10.5 & 127.4 & 38.4 & & 06:07:47.870 & $-$06:22:57.00 & M1,H & 9.9\\
 &  &  &  &  &  & &  & 12.7 & 127.2 & 62.0 &  &&   &  & \\
\hline
\end{tabular}}}
\tablecomments{
  Note. Column 1: source name; Columns 2 - 5: the pointing center coordinates, the side length of each region for the TMRT OTF observations, and the epoch of observation; Columns 6 - 10: the fitting center positions with the velocity of peak emission, the peak flux density, and the integrated flux density; Columns 11 - 13: the positions determined by interferometric observation \citep[]{Hu2016} with references H \citep[]{Hu2016}, M \citep[]{Min2000}, and M1 \citep[]{Min2001}, Column 14: the separation between the fitting centers and interferometric positions for eight sources.}
\end{sidewaystable}

\begin{figure*}
\centering
\includegraphics[width=20cm]{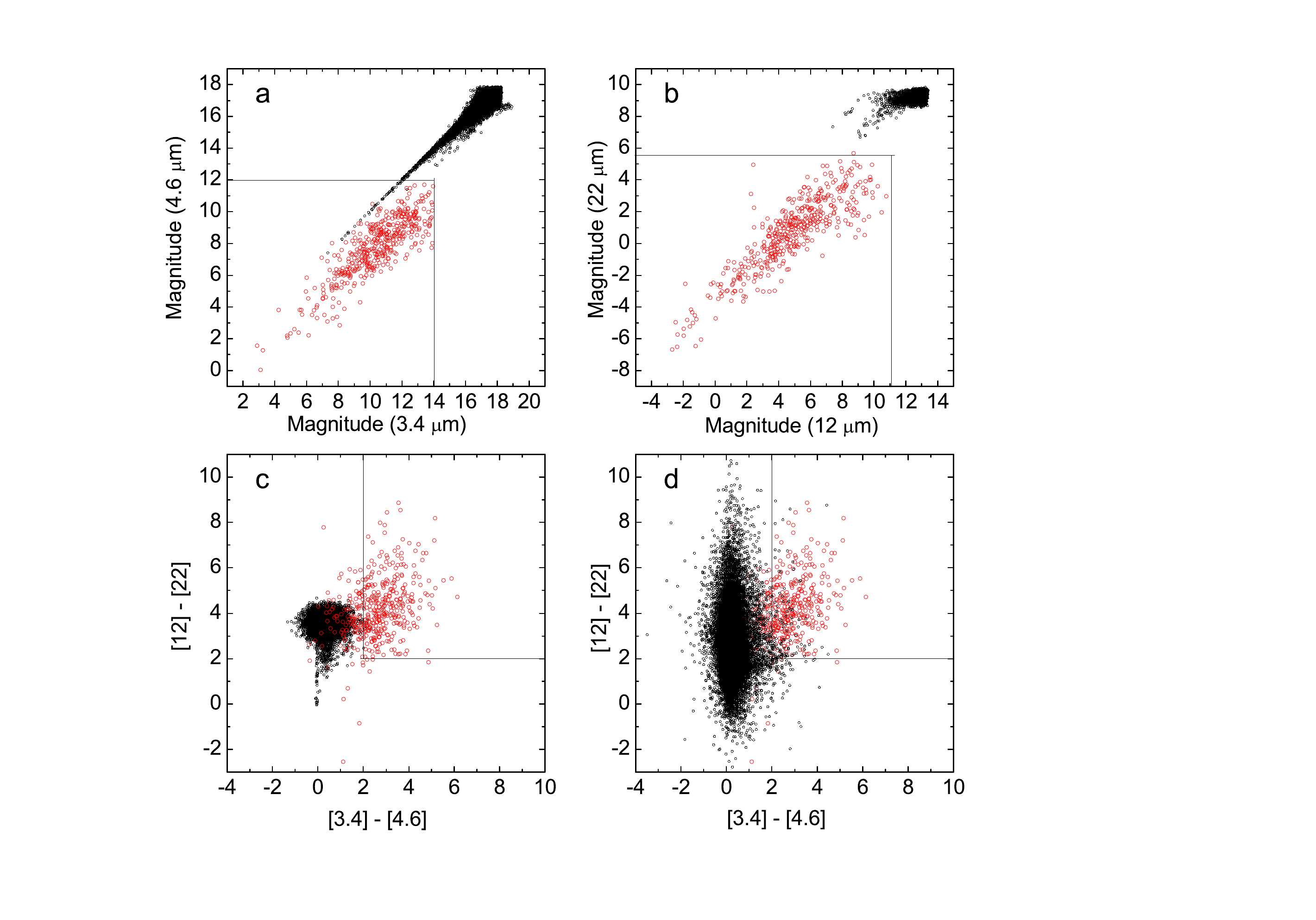}
\caption{Magnitudes and the [3.4]--[4.6] vs. [12]--[22] colors of the selected sources. The red points represent the 473 \emph{WISE} sources associated with the 6.7 GHz methanol masers from the Parkes MMB Survey. The black points represent \emph{WISE} sources from two comparison fields: G90+70 (a, b, c) and G00+00 (d). The defined box regions represent the magnitudes and color-color criteria for the \emph{WISE}-selected sources for the methanol survey in our work.}
\end{figure*}

\begin{figure*}
\centering
\includegraphics[width=13cm]{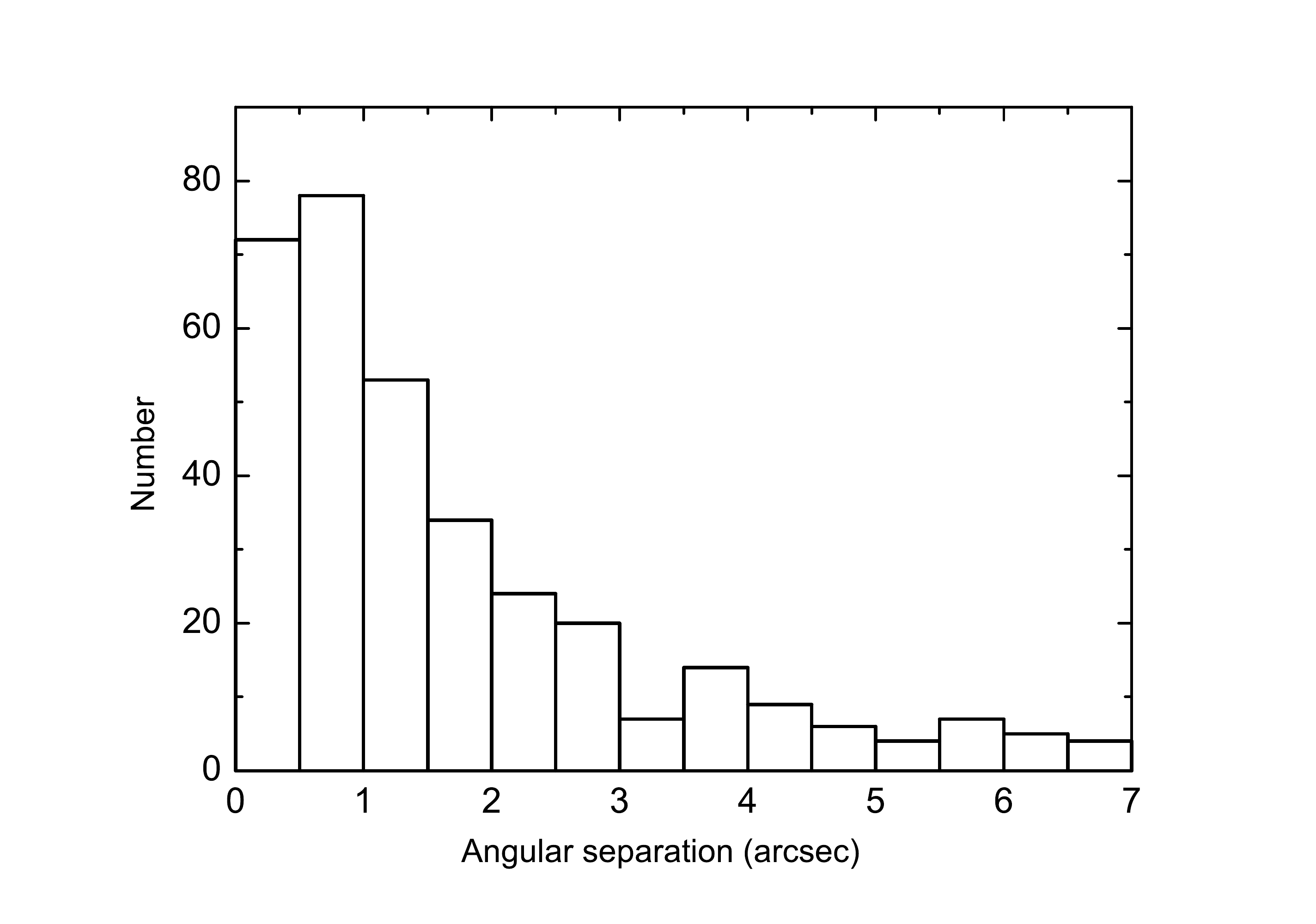}
\caption{Histogram of the distribution of the angular separation between the 330 \emph{WISE} sources and their associated MMB masers. Most of them (288 sources) have an angular separation of less than $3^{\prime\prime}$.}
\end{figure*}

\newpage
\begin{figure*}
\centering
\includegraphics[width=10cm]{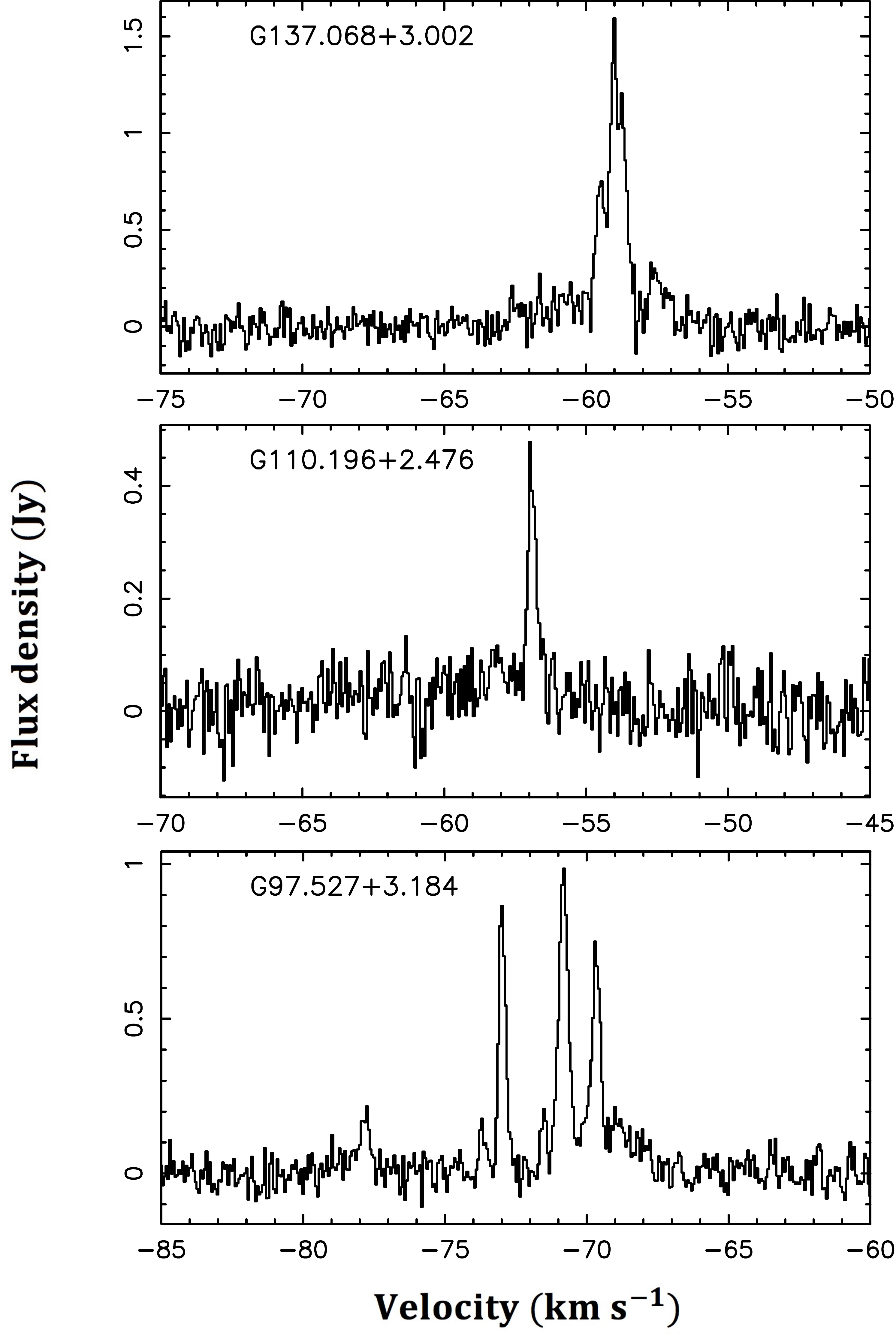}
\caption{Methanol spectra of three newly detected sources from the TMRT observations.}
\end{figure*}

\begin{sidewaysfigure}[h]
\centering
\includegraphics[width=20cm]{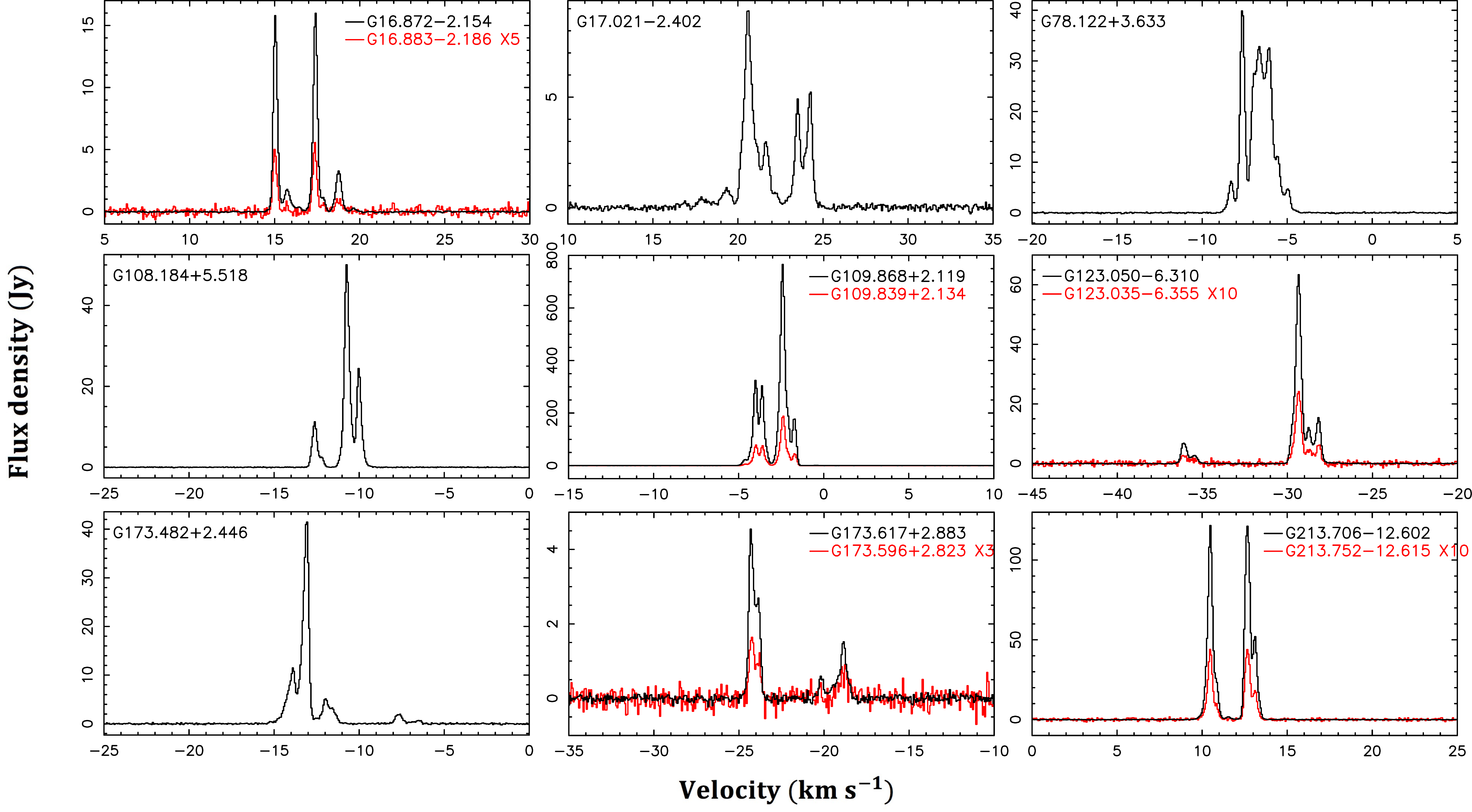}
\caption{TMRT spectra of nine previously detected methanol maser sources. For those sources having the same methanol origin, we plot their 6.7 GHz methanol spectra in the same panel with black and red lines. In order to compare the weak sources with the strong sources
in the same spot, we multiplied the flux densities of the weak sources by several times, which are written in red
words in the panel, such as G16.883$-$2.186 x5.}
\end{sidewaysfigure}

\begin{figure*}
\centering
\includegraphics[width=13cm]{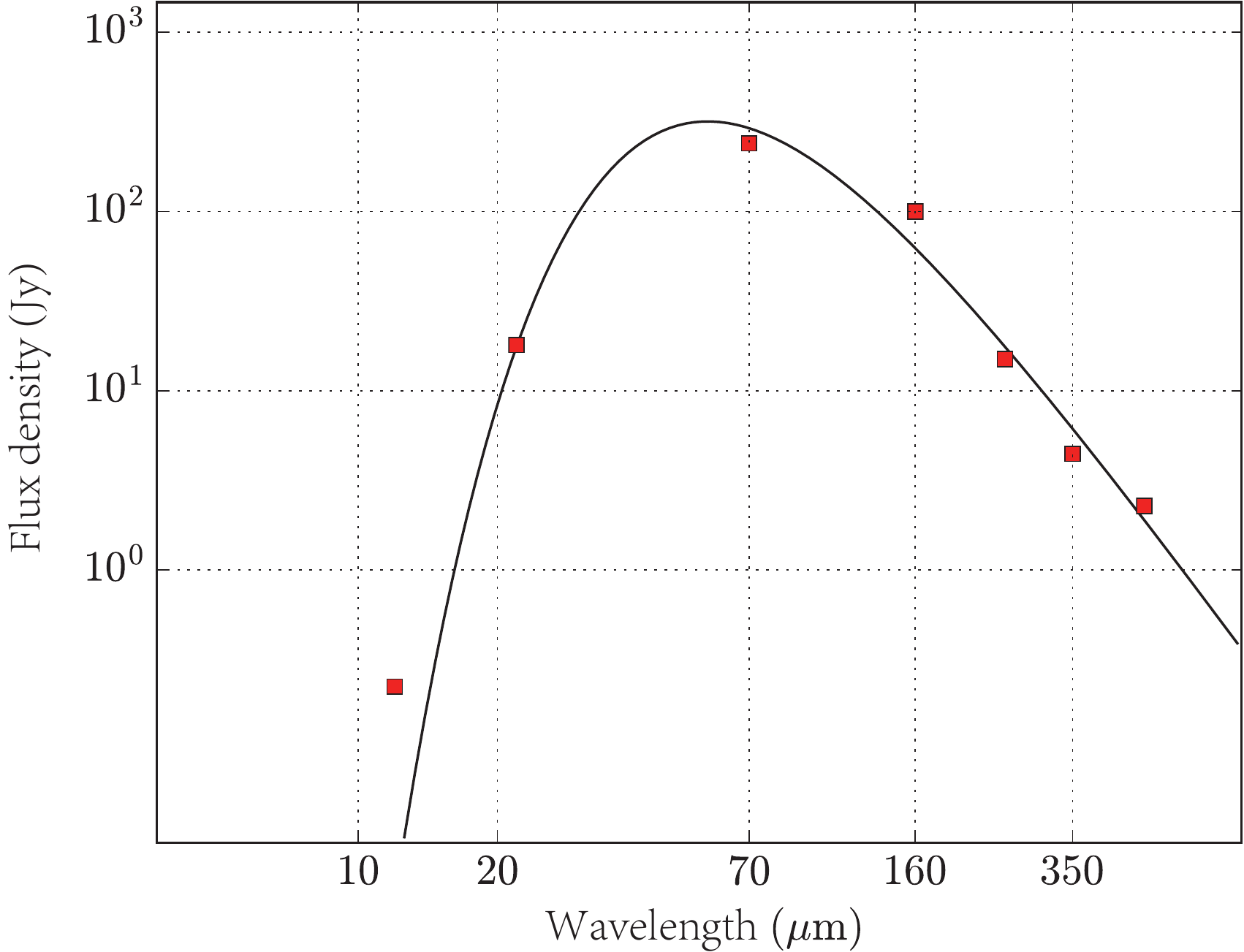}
\caption{Spectral energy distribution of G110.196+2.476. The data points at wavelengths of 12 and 22 $\mu$m are detected by the \emph{WISE}, and those at wavelengths of 70, 160, 250, 350, and 500 $\mu$m are from the \emph{Herschel} (http://irsa.ipac.caltech.edu).}
\end{figure*}

\begin{sidewaysfigure}[h]
\centering
\includegraphics[width=20cm]{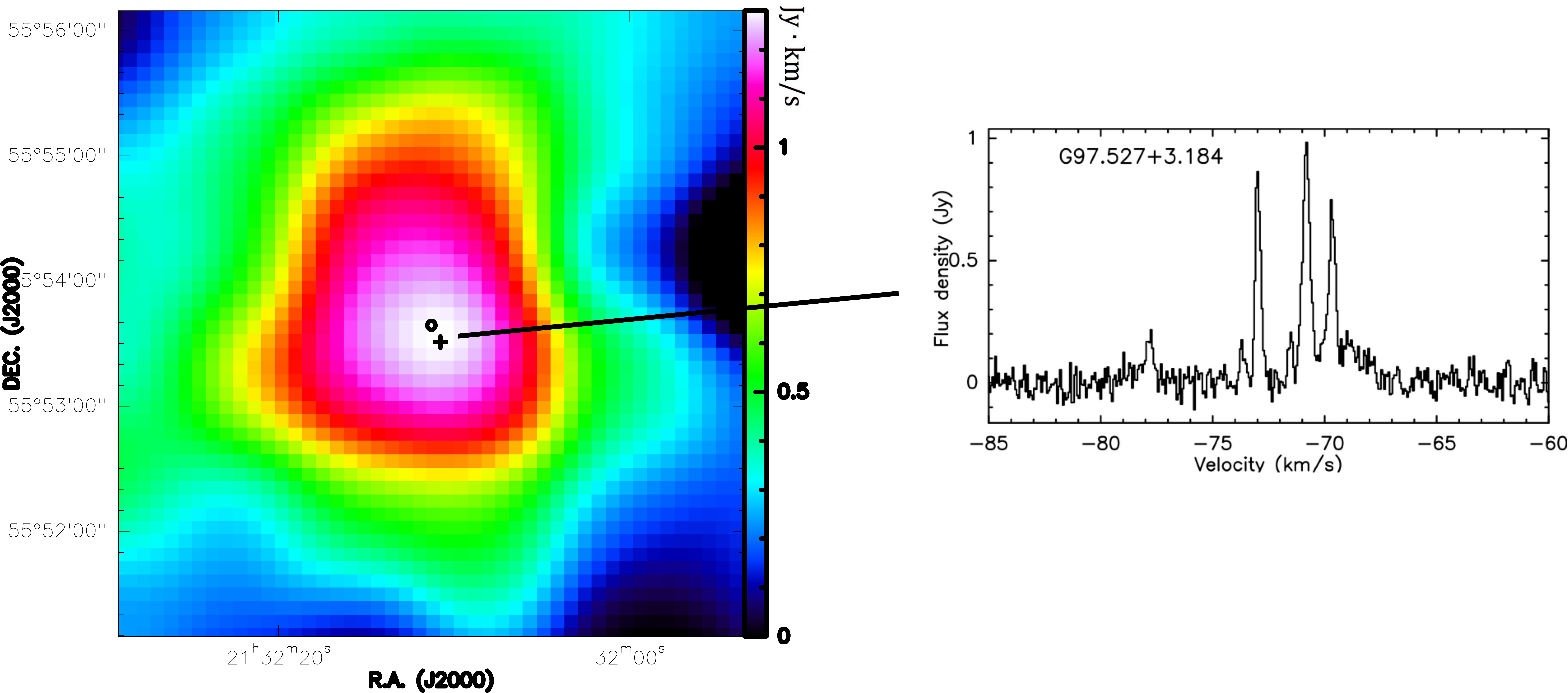}
\caption{Left: the velocity-integrated intensity map of G97.527+3.184 for the 6.7 GHz methanol maser. We derived the velocity intensity from the spectra of this source from -78.4 to $-67.6 km s^{-1}$. The ``o" represents the position of the \emph{WISE} source in the single-point-survey, and the ``+" represents the fitted peak position of the methanol emission from the TMRT OTF observation. Right: spectra in the fitting center. (The complete figures of the other 11 detected sources are available online.)}
\end{sidewaysfigure}

\begin{sidewaysfigure}[h]
\centering
\includegraphics[width=20cm]{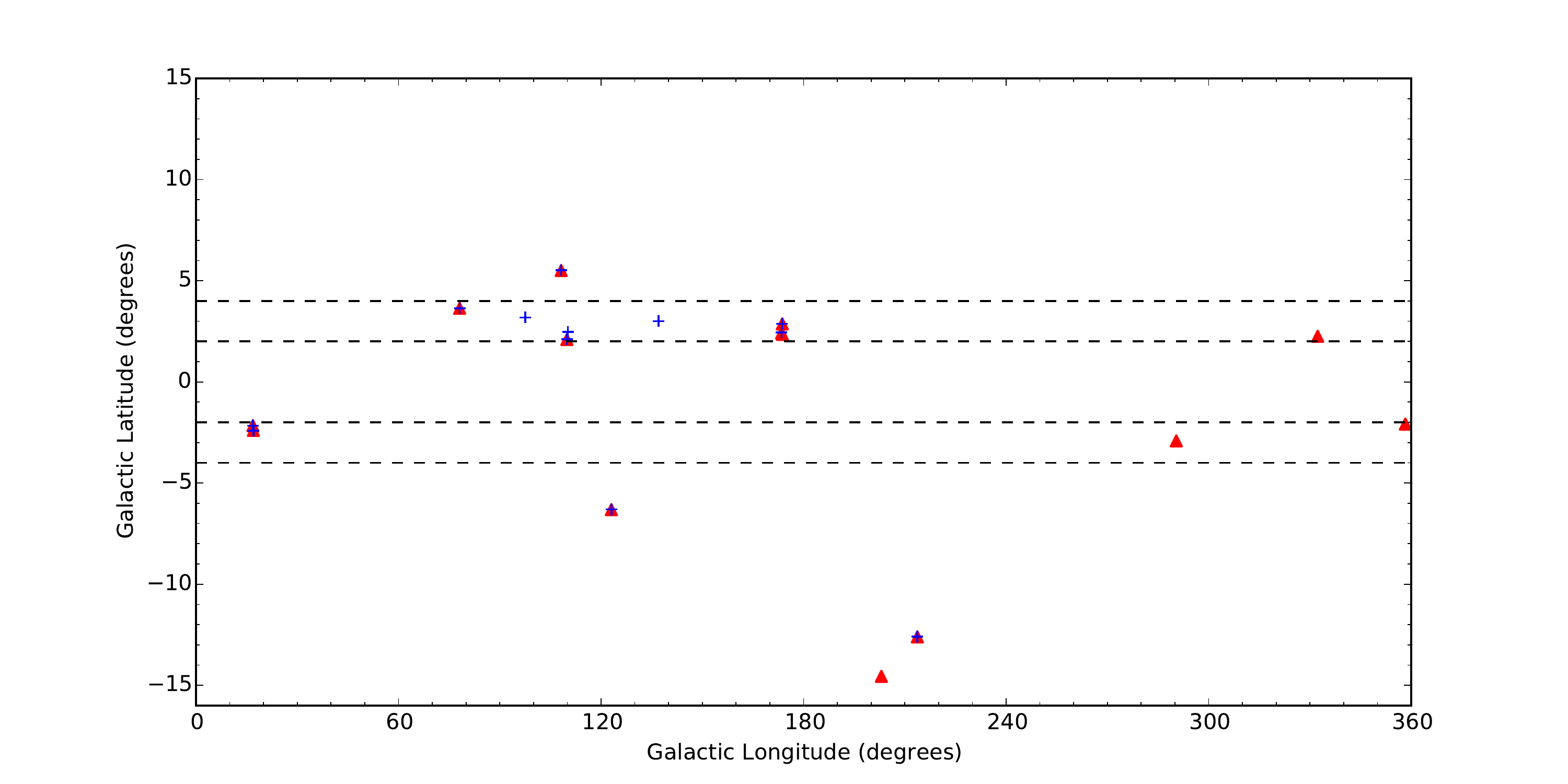}
\caption{Galactic coordinates' distribution of the 6.7 GHz methanol maser sources at high latitude, $|b|>$ 2$^\circ$. The red ``$\blacktriangle$" represents previously detected sources (Pestalozzi et al. 2005). The blue ``+" represents sources detected in our work.}
\end{sidewaysfigure}

\begin{figure*}
\centering
\includegraphics[width=15cm]{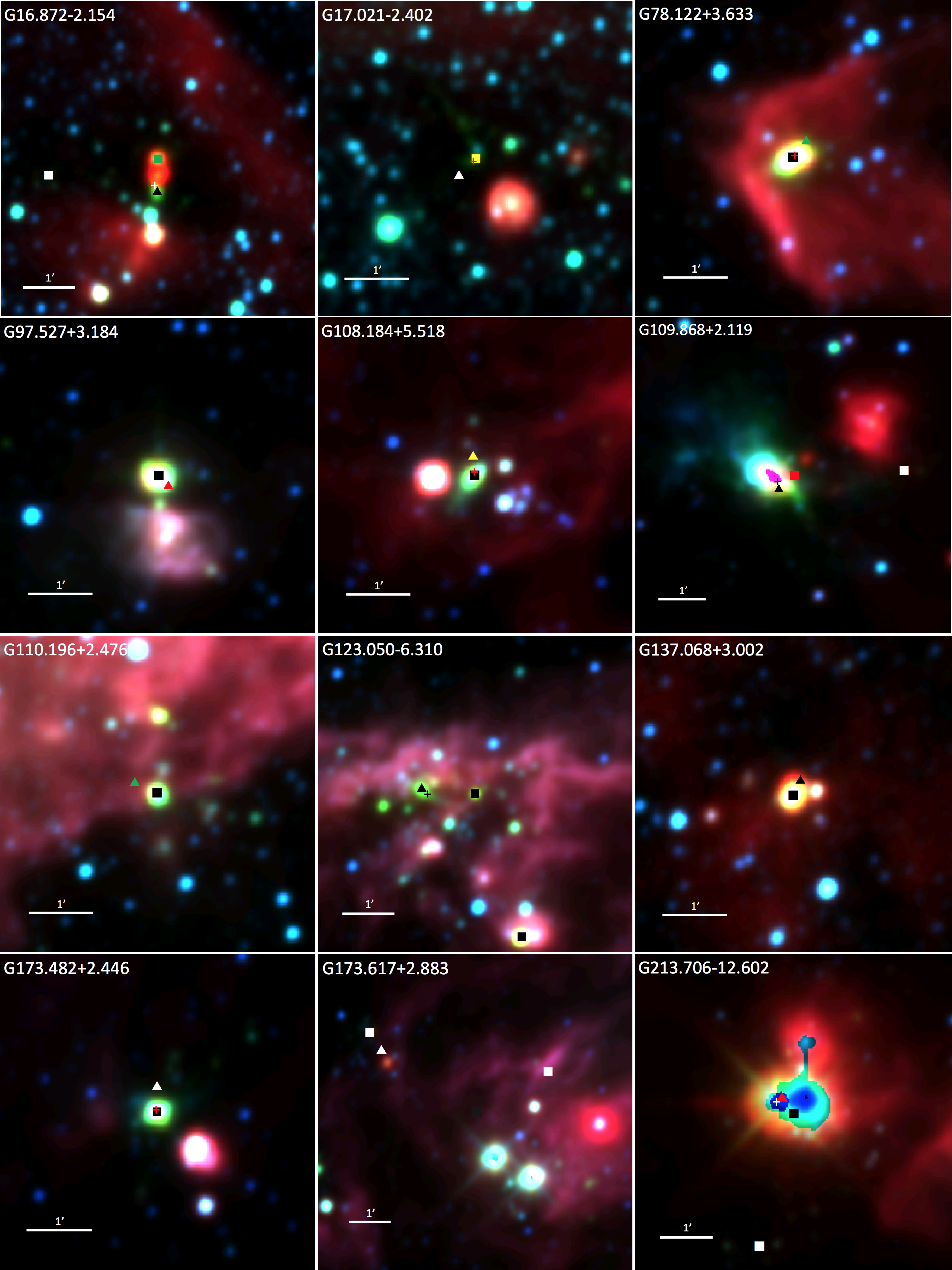}
\caption{Infrared three-color images of the 12 detected methanol sources.  The blue, green, and red colors represent 3.4, 4.6, and 12 $\mu$m bands in \emph{WISE}. The ``$\blacktriangle$" represents the fitted peak position of methanol emissions by the TMRT OTF observation (Table 3), the ``$\blacksquare$"  represents the positions of \emph{WISE} point source used in TMRT pointing observation (Table 2), and the ``+"  represents the position determined from interferometric observations (Table 3). The field of view of each region is equal to that in the OTF observation (Table 3).}
\end{figure*}

\end{document}